\documentclass{article}
\usepackage[utf8]{inputenc}

\usepackage{amsmath}
\usepackage{amsfonts}
\usepackage{float}
\usepackage{subfig}
\usepackage{graphicx, wrapfig, setspace, booktabs}
\usepackage{verbatim}

\usepackage{hyperref}
\hypersetup{
    colorlinks=true,
    linkcolor=blue,
    filecolor=magenta,      
    urlcolor=blue,
    citecolor=blue
}

\begin{document}

\title{Autocart - spatially-aware regression trees for ecological and spatial modeling}
\author{Ethan Ancell, Brennan Bean}
\date{Utah State University}

\bibliographystyle{ieeetr}

\maketitle

\section{Abstract}

Many ecological and spatial processes are complex in nature and are not accurately modeled by linear models. Regression trees promise to handle the high-order interactions that are present in ecological and spatial datasets, but fail to produce physically realistic characterizations of the underlying landscape. The ``autocart'' (autocorrelated regression trees) R package extends the functionality of previously proposed spatial regression tree methods through a spatially aware splitting function and novel adaptive inverse distance weighting method in each terminal node. The efficacy of these autocart models, including an autocart extension of random forest, is demonstrated on multiple datasets. This highlights the ability of autocart to model complex interactions between spatial variables while still providing physically realistic representations of the landscape.

\section{Introduction} \label{sec:intro}

Accurate characterizations of ecological variables have important implications for researchers and stakeholders in agriculture, watershed sciences, natural resources, and environmental sciences \cite{rodriguez2000}. Periodic land surveys to get accurate measurements for ecological variables are expensive. For example, in 2018 the United States Department of Agriculture spent over \$70 million dollars in soil surveys \cite{usda}.

Many of these ecological variables are highly interactive. As an example, soil moisture is affected by precipitation, soil composition, temperature, elevation, and the surrounding ecosystem, yet the individual relationships between each variable and soil moisture cannot be considered in isolation. The complexity of these high ordered interactions make it extremely difficult to accurately predict with traditional spatial interpolation models \cite{kathuria2019}. This is particularly true in Utah, where sharp changes in elevation create drastic changes in climate over short distances.

Traditional spatial methods such as kriging \cite{Goovaerts1997} create smooth maps of soil properties, but such approaches fail to adequately handle high-order interactions among explanatory variables. Such shortcomings emphasize the need to model interactions among the soil variables to mitigate this problem. The need for non-linear predictions of soil properties is well-documented \cite{mcbratney2003digital}. Machine-learning techniques promise to remedy under-performing linear models, due to their flexibility in characterizing complex and high-order interactions \cite{friedman2001}.

There is a lot of existing research in the realm of applying machine-learning algorithms to spatial ecological data \cite{fox2020comparing, grimm2008soil, li2011, sommer2003hierarchical, cutler2007, de2000}. Soil property mapping that utilizes machine-learning has also been extensively studied \cite{hengl2017soilgrids250m, mcbratney2003digital, ramcharan2018soil}. One particularly promising method are regression trees \cite{breiman1984}, which model high-order interactions in a way that is easy to interpret without requiring a massive amount of data like other machine-learning approaches. Unfortunately for spatial data, traditional tree-based algorithms such as regression trees have no way of accounting for the relationship between observations that is explained by their spatial distribution. Coordinate information such as longitude and latitude can be naively included as a predictor in a tree-based model, but this leads to physically unrealistic predictions, often with sharp and discrete jumps in predictions across the geographical region. An example of these visual artifacts can be seen in Figure \ref{fig:rfmap}, which shows an attempt to use the random forests algorithm to model soil moisture in Utah. This figure shows a sharp, discrete jump in the predicted soil moisture at approximately 41.5 degrees latitude. 

\begin{figure}[H]
    \centering
    \includegraphics[width = 0.5\textwidth]{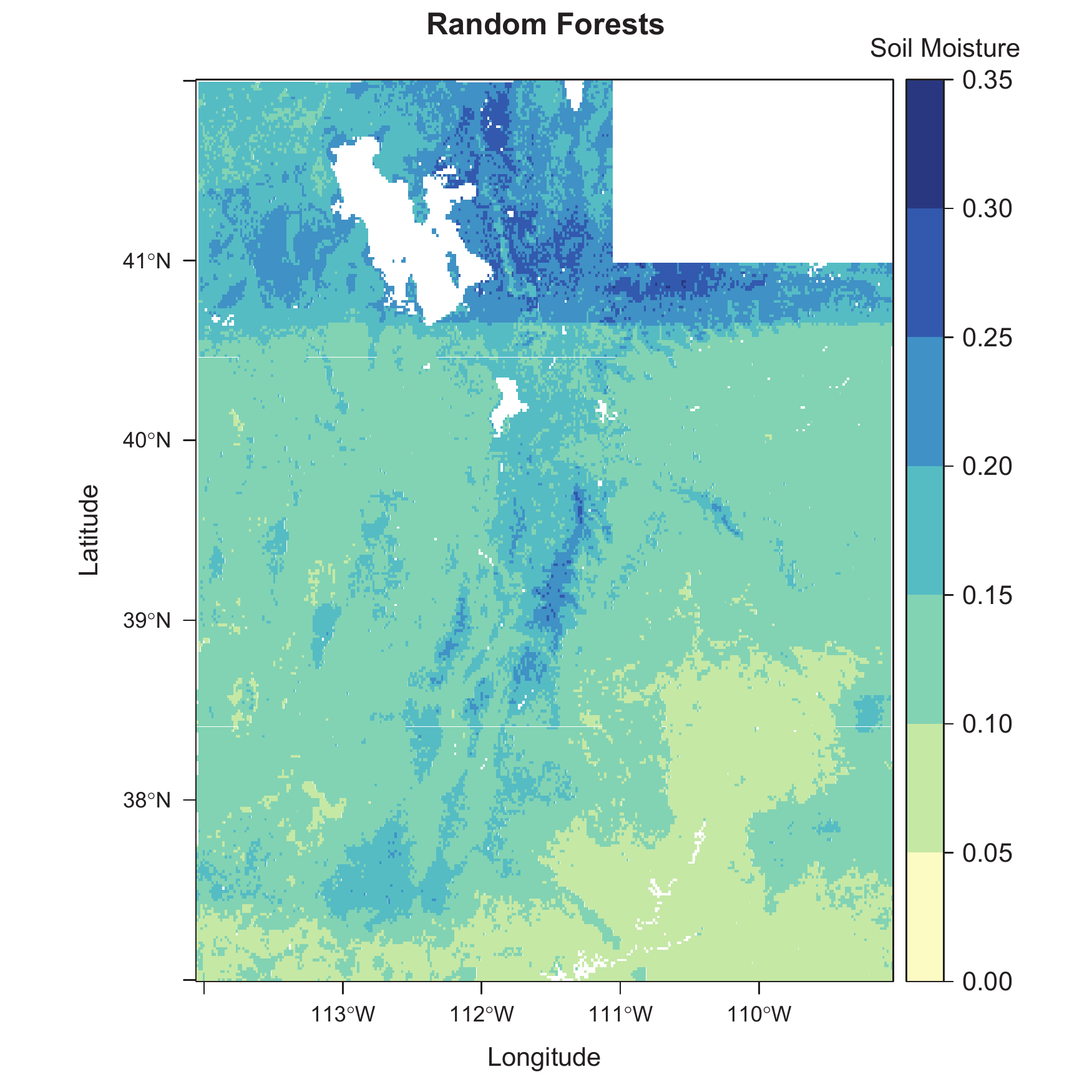}
    \caption{Prediction map of soil moisture from a default implementation of Random Forests in Utah. Explicitly including longitude/latitude as predictors in the tree-based method yields a clear jump in predictions moving from north to south that is not physically realistic.}
    \label{fig:rfmap}
\end{figure}

The visual artifacts in Figure \ref{fig:rfmap} are a symptom of the overfitting that is present in machine-learning methods when coordinates are included as explicit predictors \cite{meyer2019}. Entirely omitting coordinate information from the prediction process may remove the visual artifacts, but ignoring the spatial component can compromise accuracy. Coordinate information is especially useful when analyzing data exhibiting spatial autocorrelation, which occurs when observations are related to each other as a function of distance \cite{dubin1998}. The discussion and definitions surrounding the concept of spatial autocorrelation can be argued to directly follow from Tobler's first law of geography, which states "everything is related to everything else, but near things are more related than distant things" \cite{tobler1970}. Properly accounting for spatial autocorrelation in the modeling process is a powerful way to improve predictions in data that exhibit this property \cite{Legendre1993}. 

Another problem that exists when analyzing a spatial dataset that covers a large region is the inability to make the assumption that the distribution of the variable of interest remains consistent across the entire space \cite{kim2005, heaton2017}. Regression trees as a means to decompose a global spatial process into smaller local regions have been studied, including the effort by \cite{heaton2017} which discusses the use of hierarchical clustering trees for this purpose.

This paper seeks to extend a machine-learning algorithm to handle coordinate information in an appropriate way, avoiding the problems of overfitting and visual artifacts while fully economizing on the predictive power that resides in coordinate information. Initial attempts to create models of this type include regression tree extensions by \cite{stojanova2013, appice2014}, and a random forest extension by \cite{georganos2019geographical}. In the ensuing sections, we propose a tree-based algorithm called ``autocart'' (autocorrelative regression trees) intended to decompose a global spatial process into local regions in a hierarchical and automatic way, building upon the work proposed by \cite{stojanova2013} and \cite{appice2014}. The terminal nodes of the tree can be used for the simulation of a local spatial process using inverse-distance weighting interpolation. The result is a predictive process that harnesses the predictive power of both interpolation and regression trees.

\section{Existing work}

\subsection{Classification and regression trees} \label{sec:cart}

In the traditional regression tree algorithm, we create partition rules on a dataset to predict some response variable according to a set of splits on the predictor variables \cite{breiman1984}. The regression tree algorithm falls under the paradigm of supervised learning, where we use labeled training data to form rules for the prediction of new unlabeled training data. 

Formally, we predict a class or response $Y$ from predictor variables $X_1, X_2, \dots, X_p$ by growing a binary tree. We form a prediction with the grown tree by applying a test on one of the predictor variables at each internal node of the tree, and depending on the outcome of the test, we move to either the right or left child of the internal node and proceed to apply the next test to one of our predictor variables. The final prediction for an observation with predictors $X_1, X_2, \dots, X_p$ is made upon arriving at a terminal node of the binary tree, using the average of the response variable of the training observations that were a part of the terminal node during training.

\begin{figure}[H]
    \centering
    \includegraphics[width = 0.70\textwidth]{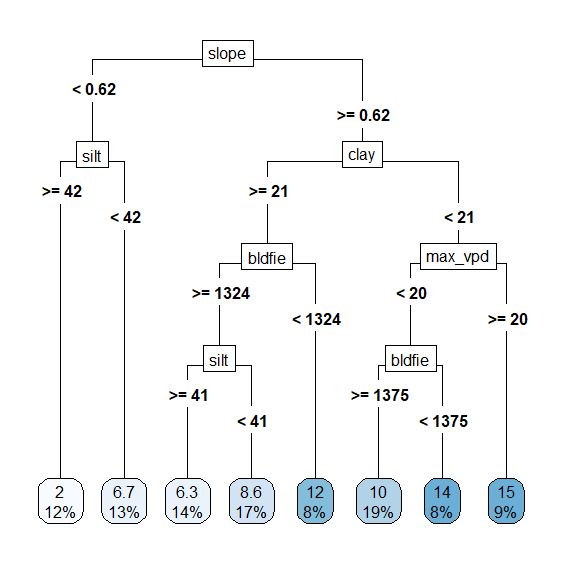}
    \caption{An example of a regression tree. The response variable for this tree is the percentage of water by volume found in a particular soil sample.}
    \label{fig:rfmap}
\end{figure}

To make a prediction, a set of predictors $\{X_1, X_2, \dots, X_p\}$ are passed into the tree. As an example, if we had predictors $\text{slope} = 0.54$ and $\text{silt} = 32$, then we would make a prediction by starting at the top node of the tree, going left because $\text{slope} = 0.54 < 0.62$, and then going right because $\text{silt} < 42$. We would arrive at the terminal node, and then make the prediction $y=6.7$. The 13\% in the terminal node indicates that 13\% of the training data for the building of the tree resides in that node.

While growing the tree, the algorithm searches for the decision rule $X_p < x$ out of all possible decision rules, such that predictive accuracy is maximized on the training dataset. This partitioning is done recursively, such that each two child nodes that are created as part of a split form the basis for the next splitting rule. To maximize predictive accuracy on the training data at each step of the algorithm, the decision rule splits the data into the two halves that minimize the residual sum of squares (RSS) in the newly formed children nodes.
\[RSS = \sum_i (y_i - \bar{y})^2\]
where $y_i$ are the observations of the labeled response value at this node, and $\bar{y} = \frac{1}{n} \sum_i {y_i}$ is the average of the labeled response variables within the node.

The minimization of $RSS$ described above can be most efficiently calculated by maximizing the variance between the nodes. This variance is calculated as
\[SS_B = n_l (\bar{y}_l - \bar{y})^2 + n_r (\bar{y}_r - \bar{y})^2\]
where $n_l$ and $n_r$ are the sample sizes in the ``left'' and ``right'' partitions respectively, $\bar{y}_l$ and $\bar{y}_r$ are the average values of the response in the left and right children, and $\bar{y}$ is the average of the response over all observations in the parent node where the split is being made.

$SS_B$ is compared to the total variance of the node if no split had been made:

\[SS_T = \sum_i (y_i - \bar{y})^2\]

The tree makes splits by maximizing the ratio between $SS_B$ and $SS_T$. The ratio of $SS_B$ and $SS_T$ is encapsulated in the so-called ``objective function'', the measure of goodness or utility of each potential split.

\begin{equation}
g_{r s s} = SS_B / SS_T
\label{eq:g_rss}
\end{equation}

\subsection{Spatial extensions to regression trees} \label{sec:existing_work}

Breiman's CART algorithm discussed in section \ref{sec:cart} is aspatial, meaning that it does not consider the geographic distribution of the measurements. As discussed in section \ref{sec:intro}, coordinate information can be included as predictor variables in the model, but often leads to physically unrealistic predictions as seen in Figure \ref{fig:rfmap}. Including coordinate information as explicit predictors in machine-learning models for spatial data also leads to an overfitting of the training dataset \cite{meyer2019}.

Simply excluding coordinate information from the predictive process may reduce the overfitting of the training dataset, but this may curtail the predictive power that lies in coordinate information, especially in spatial datasets with spatial autocorrelation.

An appropriate handling of coordinate information is to prevent the tree from making splits based upon coordinate information, but allow the tree to track coordinate information for each sample to use as part of the predictive process. The following techniques assume that the coordinate vector $\mathbf{s}=(x,y)$ is available for all samples in the training dataset. This restructure frames the regression tree prediction process as being fueled by both the information encapsulated in the predictor variables $\{X_1, X_2, \dots, X_p\}$ as well as the geographic location expressed as a coordinate vector $(x,y)$.

An extension to the regression tree algorithm was proposed by \cite{stojanova2013}. In this extension, the objective function in the regression tree algorithm is formed by a linear combination of the objective function $g_{r s s}$ described in Equation \ref{eq:g_rss} and another objective function $g_{a c}$ that optimizes for measures of spatial autocorrelation in the partitions. It is defined as 
\begin{equation}
g = (1-\alpha) g_{r s s} + \alpha g_{a c}, \quad \quad \alpha \in [0, 1]
\label{eq:stojanova_goodness}
\end{equation}
where $\alpha$ is a user-set parameter that weights minimizing the RSS versus maximizing the autocorrelative statistic. Please note that all cited equations have been converted into a common notation for convenience.

A popular statistic intended to measure the presence of spatial autocorrelation in a group of observations is Global Moran's I (i.e, Moran's I) \cite{moran1950}. Moran's I requires a vector of the response variable $Y$ where we measure spatial autocorrelation. We also require a spatial weights matrix that reflects the intensity of the spatial relationship between all pairwise observations in the group. Moran's I of a response variable $Y$ for a group of observations $G$ is defined to be
\begin{equation}
I_{Y} = \dfrac{n\sum_i \sum_j w_{i j}(y_i - \bar{y})(y_j - \bar{y})}{W\sum_i (y_i - \bar{y})^2}
\label{eq:morani}
\end{equation}
where $n$ is the total number of observations that are indexed by $i$ and $j$ in the group $G$; $w_{i j}$ is the spatial weights matrix entry that represents the weight between observations $i$ and $j$ ($w_{i j}$ is 0 when $i=j$), and $W = \sum_i \sum_j w_{i j}$. $y_i$ is the response variable of interest at observation $i$, and $\bar{y}=\frac{1}{n} \sum_i {y_i}$ is the mean of the response variable in the group of observations $G$.

Under the null hypothesis of no spatial autocorrelation, the expected value of Moran's I for the response $Y$ in group $G$ is
\[E(I_{Y}) = \dfrac{-1}{n-1}.\]
The statistic $I$ typically lies in the range $[-1, 1]$. Values of $I$ that are significantly above $E(I)$ indicate positive spatial autocorrelation, where values significantly below $E(I)$ indicate negative spatial autocorrelation \cite{dubin1998}.

A critical choice is the spatial weighting scheme to produce the spatial weights matrix entries $w_{i j}$ used in Moran's I. The following Gaussian similarity measure connecting observations $i$ and $j$ is used by \cite{stojanova2013}.
\begin{equation}
w_{i j} = 
\begin{cases}
    e^{-\frac{\text{dist}(i, j)^2}{b^2}}, & \text{if dist}(i, j) < b \\
    0, & \text{if dist}(i, j) \ge b
\end{cases}
\label{eq:autocart-weights-gaussian}
\end{equation}
where $\text{dist}( \cdot )$ is a distance metric between two observations $i$ and $j$, and $b$ is a spatial-bandwidth parameter that reflects the spatial distance at which no spatial influence is assumed between observations. Other methods for choosing $w_{i j}$ in the weights matrix for the calculation of Moran's I are viable, and the best scheme for picking $w_{i j}$ may be dataset dependent. For a typical use case, this Gaussian weighting scheme is sufficient.

In order to simultaneously maximize $I_Y$ in both partitions of the data, a fair weighting of $I_Y$ in each half according to the number of observations is required. This can be expressed as
\begin{equation}
    \Tilde{I}_Y = \dfrac{n_L \cdot I_{YL} + n_R \cdot I_{YR}}{n}
\label{eq:partition_weighted_morani}
\end{equation}
where $n_L$ and $n_R$ represent then number of observations in arbitrary ``left'' and ``right'' partitions. $I_{YL}$ and $I_{YR}$ are the value of the Moran's I statistic from equation \ref{eq:morani} for the individual left and right partitions, and $n=n_L + n_R$ is the total number of observations in the node being split.

The nature of $\Tilde{I}_Y \in [-1, 1]$ requires a re-scaling to $[0, 1]$ to ensure an even match against the residual sum of squares objective function $g_{r s s}$. The autocorrelative objective function can be defined as
\[g_{a c} = \dfrac{\Tilde{I}_Y + 1}{2}\]
which finds its way into final weighted objective function from equation \ref{eq:stojanova_goodness}:
\[g = (1-\alpha) g_{r s s} + \alpha g_{a c}, \quad \quad \alpha \in [0, 1].\]

By weighting $g_{a c}$ higher with $\alpha$, the tree is more likely to choose splits that create partitions where the observations in the partition exhibit the property of spatial autocorrelation. This is in contrast to traditional regression trees which simply seek to minimize the residual sum of squares. The motivation underlying this spatial adaptation is that the consideration of $g_{a c}$ at the expense of $g_{r s s}$ is worth the investment: the creation of spatial partitions that reflect self-contained units exhibiting a spatial pattern. \cite{stojanova2013} showed that for some datasets, weighting $g_{a c}$ higher resulted in gains in predictive accuracy with cross-validation.

On the other hand, weighting $g_{a c}$ too high has its own problems. If $g_{r s s}$ is not weighted enough, then the prediction that is formed by the mean of the response value in the terminal node $\bar{y}_T$ is not representative of the region as a whole and leads to poor predictions. The best strategy is to pick the value of $\alpha$ such that an optimal balance is struck between creating partitions that exhibit spatial autocorrelation while still maintaining the power of $\bar{y}_T$ as a baseline prediction.

\cite{appice2014} builds upon the work in \cite{stojanova2013} by proposing a novel data mining framework for geophysical data called \textit{interpolative clustering}, in which interpolation of the response variable of the training data can be used to supplement the prediction $\bar{y}_T$ of a regression tree.

As discussed in Section \ref{sec:intro}, the end goal of these modified regression trees is to decompose the global spatial landscape into focused sub-regions. \cite{appice2014} observed that the predictive power of the tree can be improved by simulating a local spatial process contained in a terminal node of the tree using an interpolative method such as inverse-distance weighting. Previously, we formed a prediction for a new observation by considering the arithmetic mean of the response variable at the terminal node of the tree $\bar{y}_T$. Now, we supplement that prediction with an interpolative simulation of the spatial pattern of the training observations in the terminal node.

In more formal terms, to make a prediction we run the predictor variables $\{X_1, X_2, \dots, X_p\}$ through the splitting rules of the tree as normal, and note the resulting terminal node $T$ that the prediction falls into. Next, we make the prediction using an interpolation method, using only the training data in the terminal node as the reference points for interpolation. We denote the coordinates of the new prediction with $\mathbf{s}=(x,y)$, whose final prediction is made with
\begin{equation}
\hat{Y}(\mathbf{s}) =
\begin{cases}
    \dfrac{\sum_{i=1}^{n_T} w_i (\mathbf{s}_{Ti}) y_{Ti}}{\sum_{i=1}^{n_T} w_i (\mathbf{s})}, & \text{if } \text{dist}(\mathbf{s}, \mathbf{s_{Ti}}) \ne 0 \text{ for all } i \\
    y_i, & \text{if } \text{dist}(\mathbf{s}, \mathbf{s_i}) = 0 \text{ for some } i
\end{cases}
\label{eq:idw_prediction}
\end{equation}
where $n_T$ is the total number of training observations in the terminal node $T$, $y_{Ti}$ are the labeled response values of the training data in $T$, $\mathbf{s}_{Ti}=(x_i, y_i)$ are the coordinate locations of the training data in $T$, and $\text{dist}(\cdot)$ is some spatial distance metric.

$w_i$ is the spatial weight that is assigned between the interpolated location $\mathbf{s}$ and the known response locations $\mathbf{s}_{Ti}$. $w_i$ is calculated with inverse distance weighting:
\begin{equation}
w_i(\mathbf{s}) = \dfrac{1}{\text{dist}(\mathbf{s}, \mathbf{s}_{Ti})^p}
\label{eq:idw}
\end{equation}
where $\mathbf{s}$ is the interpolated point, $\mathbf{s}_{Ti}$ are the known points in $T$, $\text{dist}(\cdot)$ is a distance metric from the known point to the interpolated point  (this may be Euclidean distance in the case of projected data or also great circle distance in the case of latitude/longitude coordinates), and $p \in \mathbb{R}^{+}$ is the power parameter set by the user. A higher power parameter will highly weight close observations in relation to further observations, compared to a lower power parameter where further observations have more influence in the final prediction $\hat{Y}(\mathbf{s})$. The optimal choice for $p$ is a function of the strength of the underlying spatial distribution in nature, but $p=2$ is commonly used.

The interpolative step described in equations \ref{eq:idw_prediction} and \ref{eq:idw} is most effective when used in combination with the objective function $g_{a c}$. Creating partitions and terminal nodes that exhibit spatial autocorrelation is beneficial for inverse-distance weighting interpolation as its efficacy relies upon the assumption that the correlation between observations decreases as the distance between them increases.

\section{Autocart: extensions to the spatial regression tree} \label{sec:ranged_interpolation}

In this section we propose further extensions to the methodologies introduced in Section \ref{sec:existing_work}. We refer to the final tree algorithm as the ``autocart'' algorithm (autocorrelative regression trees), which is publicly available as an R statistical software package at the following URL:
\begin{center}
    \href{https://github.com/ethanancell/autocart}{github.com/ethanancell/autocart}
\end{center}

\subsection{An adaptive approach to picking the power parameter $p$}

In Section \ref{sec:existing_work}, inverse distance weighting at the terminal nodes of the regression tree is discussed as a way to supplement the prediction of the tree. Here, we propose a novel approach to picking the power parameter $p$ in equation \ref{eq:idw} for the inverse-distance weights.

If we consider these spatial regression trees to do the part of automatically decomposing a global landscape into local areas where the spatial pattern may differ, then it is unfair to state that the optimal choice of the power parameter $p$ is constant across all terminal nodes representing local regions. A local region may exhibit a stronger dependence between closely neighboring observations than other local regions, necessitating a varying $p$.

In order to assess the strength of a spatial relationship in some terminal node $T$, we can reuse the Moran's I statistic from equation \ref{eq:morani}. A terminal node exhibiting a stronger correlation between closely neighboring observations will have a comparatively higher value of $I_Y$. In a region where a stronger correlation is noted between closely neighboring observations, it is sensible to pick $p$ to be higher so that the weights defined in equation \ref{eq:idw} give greater weight to neighboring observations as compared to far observations. In a region where a weak correlation is noted between closely neighboring observations, it is sensible to pick $p$ to be lower so that the resulting prediction catches on to the trend of the region as a whole, rather than relying upon inappropriate confidence in the predictive power of close observations, indicated by the low value of $I_Y$.

Let $M$ represent the set of $I_{Yi}$ calculated for each terminal node in the regression tree. In the case that $I_{Yi} < E(I_{Yi})$ for the terminal node, there is no value in interpolation as the correlation between observations close in space is not significant. The ranged value of $p$ will be only based upon the values $I_{Yi} \in M$ where spatial autocorrelation is observed. We denote this new set with
\[\Tilde{M} = \{I_{Yi} \in M : I_{Yi} > E(I_{Yi})\}.\]

To make a prediction for the coordinate $\mathbf{s} = (x,y)$, we first run the accompanying predictor variables $\{X_1, X_2, \dots, X_p\}$ through the tree to find which terminal node $\mathbf{s}$ belongs to. Once we have found the correct terminal node $T$, we make the prediction in a similar way to equation \ref{eq:idw_prediction}:
\begin{equation}
    \hat{Y}(\mathbf{s}) =
    \begin{cases}
        y_i, & \text{if dist}(\mathbf{s}, \mathbf{s}_i) = 0 \text{ for some } i \\
    
        \bar{y}_T, & \text{if } I_T < E(I_T) \text{ and dist}(\mathbf{s}, \mathbf{s}_i) \ne 0 \text{ for all } i \\
        
        \dfrac{\sum_{i=1}^{n_T} w_i(\mathbf{s})y_i}{\sum_{i=1}^{n_T} w_i(\mathbf{s})}, & \text{if } I_T > E(I_T) \text{ and dist}(\mathbf{s}, \mathbf{s}_i) \ne 0 \text{ for all } i
    \end{cases}
\label{eq:idw_prediction_ranged}
\end{equation}
where $I_{YT}$ represent the observed value of the Moran's I statistic for the training observations in the terminal node $T$ with respect to the response variable $Y$, and $E(I_{YT})$ is the expected value of $I_{YT}$. $\bar{y}_T$ denotes the observed mean of the response variable $Y$ of the training observations in $T$.

$w_i$ remains much the same as in equation \ref{eq:idw}, the difference being we use the varying power parameter $p_T$:
\begin{equation}
    w_i(\mathbf{s}) = \dfrac{1}{\text{dist}(\mathbf{s}, \mathbf{s}_i)^{p_T}}.
\end{equation}
The varying power parameter $p_T$ is a monotonic increasing function that maps from $[\min(\Tilde{M}), \max(\Tilde{M})]$ to $[p_1, p_2]$, where $p_1$ and $p_2$ are user-set parameters that indicate the range of power parameters. The terminal node that exhibits the most significant value of $I_{YT}$ will use $p_2$ for its power parameter, and the terminal node with the least significant (yet above expected) value of $I_{YT}$ will use $p_1$.

One choice of $p_T: [\min(\Tilde{M}), \max(\Tilde{M})] \mapsto [p_1, p_2]$ is the following linear function:
\begin{equation}
    p_T = \dfrac{(I_T - \min(\Tilde{M}))(p_2 - p_1)}{\max(\Tilde{M}) - \min(\Tilde{M})} + p_1 
\end{equation}
As $p_1$ and $p_2$ are set by the user, it is crucial that the user have a sense of an appropriate range of values for $p$ in the context of their particular dataset. 

\subsection{An objective function for the encouragement of spatially-compact terminal nodes} \label{sec:beta}

Section \ref{sec:existing_work} described an objective function that encourages high values of spatial autocorrelation within the internal nodes of the tree. Section \ref{sec:existing_work} also describes the use of interpolation at the terminal nodes of the tree to supplement the prediction of $\bar{y}_T$. In this section, we propose another possible objective function for the tree and explore the possibility of weighting this new objective function alongside the objective function $g_{a c}$ described in section \ref{sec:existing_work}.

When using interpolation as part of the predictive process, it is desired that terminal nodes of the tree create sub-regions of the data that are ideal for interpolation. Excessive overlap in the regions create by the terminal nodes is not ideal for interpolation, as much of the final prediction is weighted by distant observations while ignoring other observations that are geographically close but not in the same terminal node. In this section, another objective function $g_{s c}$ for the encouragement of spatially-compact internal and terminal nodes is introduced.

At an arbitrary level of the splitting process, define the total sum of squared pairwise distances within the node $N$ to be
\[TSS_D = \sum_{\mathbf{s}_i \in N} \sum_{\mathbf{s}_j \in N} \text{dist}(\mathbf{s}_i, \mathbf{s}_j)^2.\]
Consider an arbitrary partition of the data in the node $N$ that produces a ``left'' and ``right'' partition, sub-scripted by $l$ and $r$ respectively. Let $N_l$ be the set of all training observations in the left partition, and $N_r$ be the set of all training observations in the right partition.

Define the between sum of squared differences for the partitions to be
\[BSS_D = \sum_{\mathbf{s}_i \in N_l} \sum_{\mathbf{s}_j \in N_r} \text{dist}(\mathbf{s}_i, \mathbf{s}_j)^2.\]
As a sort of ``spatial extension'' to a 1-way anova, the total sum of squares of distances is composed of the sum of all between sums of squared differences and the sum of all within sums of squared differences. This is represented as
\begin{equation}
    TSS_D = BSS_D + WSS_D
\label{eq:anova_distances}
\end{equation}
where
\[WSS_D = \sum_{\mathbf{s}_i \in N_l} \sum_{\mathbf{s}_j \in N_l} \text{dist}(\mathbf{s}_i, \mathbf{s}_j)^2 + \sum_{\mathbf{s_i} \in N_r} \sum_{\mathbf{s}_j \in N_r} \text{dist}(\mathbf{s}_i, \mathbf{s}_j)^2.\]

Minimizing $WSS_D$ encourages spatially compact regions resulting from the split, minimizing uncertainty in the interpolation step. Due to the identity in equation \ref{eq:anova_distances}, this is possible by maximizing $BSS_D$.
The previous objective functions $g_{r s s}$ and $g_{a c}$ from sections \ref{sec:cart} and \ref{sec:existing_work} respectively indicate a more desirable split when $g_{r s s}$ and $g_{a c}$ are higher in value and closer to 1. As $\dfrac{BSS_D}{TSS_D} \in [0,1]$, it is natural and intuitive to define the goodness of spatial compactness $g_{s c}$ as:

\begin{equation}
    g_{s c} = \dfrac{BSS_D}{TSS_D}.
\label{eq:gsc}
\end{equation}

We can include $g_{s c}$ in the linear combination of previously discussed objective functions with the weighting parameter $\beta$:

\begin{equation}
g = (1-\alpha-\beta)g_{r s s} + \alpha g_{a c} + \beta g_{s c}, \quad \quad \text{where } \alpha, \beta \in [0, 1] \text{ and } \alpha+\beta \le 1.
\label{eq:final_goodness}
\end{equation}

Thus, the revised regression tree algorithm is to search through all predictor values $X_1, X_2, \dots, X_p$ looking for the splitting rule $x < X_i$ for some $i$ that maximizes the objective function $g$ of equation \ref{eq:final_goodness} at each recursive partitioning of the data.

\section{Autoforest - a Random Forest extension to autocart trees}

The revised objective function in equation \ref{eq:final_goodness} and the interpolative process discussed in section \ref{sec:ranged_interpolation} are promising ways to improve upon the predictions of regression trees when applied to continuous spatial data. Random Forests are a powerful extension of classification and regression trees. Random Forests increase the predictive accuracy of classification and regression trees by minimizing the variance of predictions by producing a ``forest'' of trees, each trained using a bootstrapped sample of training data and a random subset of the predictor variables to split on at each node. The averaging of predictions that occurs in Random Forests greatly improve upon the predictive power of a single regression tree \cite{breiman2001random}.

The creation of a ``forest'' of autocart trees is proposed and discussed in this section.

Let us denote a single autocart tree as a function $f_A$, where a prediction is made by running a set of predictors $\{x_1, x_2, \dots, x_p\}$ through the splitting rules in the tree (trained by maximizing the objective function $g$ at each recursive partition), and then assigning the final prediction either by the average of the response variable in the terminal node (previously referred to as $\bar{y}_T$), or by an interpolative rule $\hat{Y}(\mathbf{s})$ as in equation \ref{eq:idw_prediction} or equation \ref{eq:idw_prediction_ranged}.

We create a forest of $k$ autocart trees by creating the set of trees $F = \{f_{A_1}, f_{A_2}, \dots, f_{A_k}\}$. A prediction is made by running the set of predictors $\{x_1, x_2, \dots, x_p\}$ through each tree, and then using the arithmetic mean of the prediction of all trees in $F$:

\[\hat{Y} = \dfrac{1}{k} \sum_{i=1}^{k} f_{A_i}(\{x_1, x_2, \dots, x_p\}).\]

Each regression tree $f_{A_i}$ is trained with $\dfrac{2n}{3}$ training observations randomly sampled from all $n$ observations without replacement. Note that this differs from standard Random Forests where $n$ observations are sampled with replacement. In this spatial adaption, repeat observations with identical coordinate information causes problems in the spatial weights matrix, as this results in an ``infinite'' weight.

As all $n$ records have an equal chance of being chosen, the bias of $f_{A_i}$ is not affected, especially across all $k$ trees. Additionally, in each tree only $m$ predictors are selected from ${X_1, X_2, \dots, X_p}$ at each node. $m$ is a user-set parameter, but can be safely defaulted to $\lceil \frac{p}{3} \rceil$.

The autoforest extension to the autocart tree is a way to imbue the Random Forest algorithm with spatial data while refraining from explicitly including coordinates as a predictor in the splitting.

\medskip

(\textit{Note: The software implementation of autoforest currently chooses $m$ predictor variables to split on at each tree rather than at each node. This was done for ease of implementation, but a future version of the R package will resolve this issue.})

\section{Results and comparison of model architectures} \label{sec:datasets_and_results}

\subsection{Datasets Tested}

An optimal dataset for the autocart algorithm would include coordinate information for all training observations, and the prediction of a continuous response variable over the region represented by the dataset. Additionally, all predictor variables used to train the autocart tree must be available at all locations where predictions are desired, as techniques to infer the value of a missing predictor variable $X_i$ are not covered in this paper. If the autocart algorithm is to be used in a mapping setting, then gridded data across the entire mapped region is required.

\subsubsection{September 2017 Utah soil moisture}

This dataset contains the average soil moisture level recorded in moisture per cubed centimeter for 195 remote sensing stations across the state of Utah. Gridded 30-year PRISM Climate Normals \cite{oregonstate2004} are used, including the 30-year normals for maximum vapor pressure deficit, mean annual temperature, and mean annual precipitation. Additionally, a digital elevation map of Utah from the PRISM Climate Normals is used to obtain the elevation predictor and the derived slope and aspect predictors. 

The 30-year PRISM Climate Normals and derived data are selected for their gridded nature and possible environmental relation to soil moisture.

\begin{center}
\begin{tabular}{|c|p{10cm}|}
    \hline
    \textbf{Variables} & \textbf{Description} \\
    \hline
    sm\_8 & The proportion of water per $\text{cm}^3$ of soil \\
    \hline
    elev & The elevation of the location in meters \\
    \hline
    slope & A unitless ``rise over run'' measurement of the surface angle of the location. This is calculated from the ``elev'' model. \\
    \hline
    aspect & The compass orientation of the slope at a point measured in degrees, where 0 and 360 degrees is north, 90 degrees is east, etc. \\
    \hline
    min\_vpd and max\_vpd & The 30-year estimate of minimum / maximum vapor pressure deficit measured in kilopascals (kPa) \\
    \hline
    min\_temp, max\_temp, and mean\_temp & The 30-year estimate of minimum, maximum, and mean temperature measured in degrees Fahrenheit \\
    \hline
    mean\_dewpoint & The 30-year estimate of the mean dew point temperature in degrees Fahrenheit \\
    \hline
    precip & The 30-year estimate of annual precipitation in inches \\
    \hline
\end{tabular}
\end{center}

\subsubsection{Utah 2017 Ground snowload dataset}

This dataset contains the 50 year ground snow load at a variety of measurement stations across the state of Utah \cite{Bean2018-report}. Predictors are obtained from gridded PRISM 30-year Climate Normals \cite{oregonstate2004}:

\medskip

\begin{center}
\begin{tabular}{|c|p{10cm}|}
    \hline
    \textbf{Variables} & \textbf{Description} \\
    \hline
    yr50 & The response variable: measures the designed ground snow load at the site in kilopascals (kPa) \\
    \hline
    ELEVATION & The elevation of the measurement site in meters \\
    \hline
    PPTWT & The total amount of precipitation in a year in inches\\
    \hline
    MCMT & The mean temperature of the coldest month in the year in Celsius \\
    \hline
    MWMT & The mean temperature of the warmest month in the year in Celsius \\
    \hline
\end{tabular}
\end{center}

\medskip

To fix the skewed distribution of the yr50 variable, a log transform was taken of the response.

\subsubsection{Kenya Poverty Mapping}

This dataset contains variables related to mapping the presence of poverty in various states of Kenya \cite{columbia2005, ilri2005}.

The variables in the dataset include the following:

\begin{center}
\begin{tabular}{|c|p{9 cm}|}
    \hline
    \textbf{Variables} & \textbf{Description} \\
    \hline
    POORDENS & The number of poor people per $\text{km}^2$ \\
    \hline
    AREA & The area of the active community group in Kenya \\
    \hline
    FLIVDEN & The density of livestock expressed in total livestock units/$\text{km}^2$ \\
    \hline
    ACCWAT & The percentage of area within one hour walking distance of a permanent natural water source. \\
    \hline
    PTOWNKM & Distance from the shopping center in each sublocation to the nearest major towns by road, in kms. \\
    \hline
    GROUPDENS & The total number of active community groups, including non-governmental organizations and community based organizations. \\
    \hline
    LATITUDE & The latitude of the centroid of the community group (obtained from accompanying shapefile) \\
    \hline
    LONGITUDE & The longitude of the centroid of the community group (obtained from accompanying shapefile) \\
    \hline
\end{tabular}
\end{center}

\subsection{Results} \label{sec:results}

We use cross-validation to assess the predictive accuracy of each model. In cross-validation, we divide the data into $k$ disjoint partitions known as ``folds''. The model is trained on $k-1$ folds, and then used to predict the response variable $Y$ of the withheld fold. The predictions $\hat{Y}$ of the model can be compared to the real response $Y$ for an assessment of the performance of the model. We repeat the training on $k-1$ folds of the data $k$ times, withholding a different fold each time, such that all data in the training data eventually has the chance to be withheld and compared to a prediction from the model where the fold was absent. This strategy of withholding information at each step provides an estimate of a model's ability to predict new observations and discourages over-fitting the input data. Cross-validation is the gold standard for the assessment of a model when a separate testing dataset is not available.

One choice in forming the $k$ folds is to randomly select observations from the training data to be a part of each fold. However, the autocorrelation present in spatial datasets can cause traditional cross-validation to under-estimate the true model error. \cite{meyer2019} discusses this issue and presents a solution for the cross-validation of spatial data known as ``spatial blocking''. In this setup, the folds in cross-validation are formed by creating chunks of neighboring observations, which limits the opportunity for geographically close neighbors to be a part of different folds in cross-validation.

If the spatial blocks that form the cross-validation folds are too large, then the predictive power of the model may be \textit{underestimated}, as in a realistic setting predictions may be often made at sites that are very close to the training observations. Additionally, if we consider the regression tree models discussed in this paper to be a tool for decomposing a global spatial process into separate local processes, then withholding all data in a large spatial block region may inadvertently eradicate the local spatial region's representation. On the other hand, if the spatial blocks are too small, then the number of folds may be very large and dramatically increase the computation time required to perform the cross-validation procedure. In the absence of well-defined rules regarding the geographical size of the cross validation folds, groups were constructed through the process of trial and error through a consideration of the maximum geographical distance between within-group observations. In the case of both the Utah 2017 snow and soil datasets, a distance of 15km was chosen. For the Utah 2017 snow dataset, this yields around hundred sub-groups which were then consolidated into 10 larger groups for use in cross-validation. 

Once a vector of predictions from the model has been created for all 10 folds, the results of each algorithm on each dataset are evaluated with the root mean square error (RMSE). This is a common metric to assess the predictive accuracy of cross-validation for continuous regression problems. 

\[RMSE = \sqrt{\dfrac{1}{n} \sum_{i=1}^{n} (\hat{y}_i - y_i)^2}\]

where $n$ is the number of observations in the dataset used for cross-validation, $y_i$ is the $i$th element of the true response vector of the training data, and $\hat{y}_i$ is the $i$th element of the prediction vector made by the model from 10-fold cross-validation.

\subsubsection{A tuning method for autocart}

As the autocart function contains several tune-able parameters (namely $\alpha$, $\beta$, and the spatial bandwidth '$b$'), we need to be careful how we select the optimal choice of parameters during cross-validation. The ``best'' performing choices of $\alpha$, $\beta$, and $b$ over all the training data may be different than the best performing choices for a random subset of the data, such as a subset used in cross-validation. In a realistic scenario, we predict new data that was not a part of the training data, and thus we do not have the labeled response variable $y$ to tune the parameters with. Thus, instead of tuning the parameters $\alpha$, $\beta$, and $b$ over all the data and then performing cross-validation, we tune the parameters to the 9 withheld groups, and then predict onto the last ``test'' group. In this way, the optimal choices for $\alpha$, $\beta$, and $b$ will likely vary each time we withhold a group. In the next section, the cross-validated accuracy of autocart is obtained using this tuning method.

\subsubsection{Dataset results} \label{sec:result_table}

To test the datasets, 6 different methods are used:

\begin{itemize}
    \item \textbf{``Simple'' Prediction}: This is a baseline prediction method that ensures a machine-learning method is providing an improvement over an overly-simplistic model. The simple prediction is formed by taking the average of the response variable in the nine withheld groups, and then using that average to predict for the withheld group in cross-validation.
    \item \textbf{Regression trees}: Simple regression trees that are introduced at the beginning of the paper. The trees are pruned to the point with the best cross-validated accuracy.
    \item \textbf{Autocart with $p=2$}: An autocart interpolation tree using the power parameter of $p=2$.
    \item \textbf{Autocart with $p_1 = 0.5, p_2 = 3.0$}: An autocart interpolation tree using the ranged power parameter $p_1$ and $p_2$, meant to provide a comparison to the unranged power parameter $p=2$.
    \item \textbf{Random forest with ntree = 100}: A random forest made of 100 regression trees.
    \item \textbf{Autoforest with ntree = 100}: An autoforest made up of 100 autocart trees.
\end{itemize}

In each column representing a dataset, the RMSE of the ``best performing'' model is written in bold font.

\begin{center}
\begin{tabular}{|c||p{3.5cm}|p{3.4cm}|p{3.4cm}|}
    \hline
    \multicolumn{4}{|c|}{\textbf{RMSE of spatial cross-validation}} \\
    \hline
     & \multicolumn{3}{|c|}{\textbf{Dataset}} \\
    \hline
    \textbf{Method} & September 2017 Soil (proportion of water composition per $\text{cm}^3$) & Utah 2017 Snow (log of 50 year ground snow load avg in kPa) & Kenya Poverty Mapping (log of number of poor residents per $\text{km}^2$) \\
    \hline
    ``Simple Prediction'' & 0.0882 & 0.8890 & 1.219 \\
    \hline
    Regression trees & 0.1082 & 0.3445 & 1.255 \\
    \hline
    Autocart with $p=2$ & 0.0962 & 0.3097 & 0.966 \\
    \hline
    Autocart with $p_1=0.5, p_2 = 3.0$ & 0.0935 & 0.3089 & 0.989 \\
    \hline
    Random forest with $\text{ntree} = 100$ & 0.0871 & \textbf{0.2845} & \textbf{0.933} \\
    \hline
    Autoforest with $\text{ntree} = 100$ & \textbf{0.0842} & 0.3003 & 0.993 \\
    \hline
\end{tabular}
\end{center}

\section{Discussion of Results} \label{sec:result_discussion}

\subsection{Inadequacies in the soil moisture datasets}

In the "September 2017 Soil" dataset, we observe that the simple regression using the average of the response in the nine withheld groups was nearly the best performing method, outperforming all except Random Forests by a slim margin.

This highlights an inadequacy in the data, as none of the tested machine-learning methods are capable of learning the patterns in the labeled response variable given the set of gridded predictor variables. The following are possible explanations for the poor performance of the models on the data:

\begin{enumerate}
    \item \textbf{The variation in soil moisture may be much more ``localized'' than previously thought.} As the soil moisture data is only available at 195 sites in the 2017 soil moisture dataset, there is a strong possibility that there does not exist enough data for the machine-learning models to appropriately characterize the patterns in the landscape.
    \item \textbf{The given gridded predictor variables have no relationship with soil moisture.} One requirement for candidate predictor variables is that they are available as high resolution gridded maps for the area of interest. The number of candidate variables satisfying this requirement is limited. Thus, there may be variables better suited for soil moisture prediction, but are unusable in their current forms. 
    \item \textbf{The data may be contaminated.} Some of the sampling locations are yield unusually high soil moisture measurements. This may be the result of an improper inclusion of irrigation site data (yielding an artificially high measurement of soil moisture) or perhaps an anomalous rainy day. Such anomalies defy relationships that may otherwise exist between soil moisture and the response variables, but there is no current way to know which observations should be removed due to human intervention in the soil moisture content. 
\end{enumerate}

Datasets covering other time periods were supplied by the Plants, Soils, and Climates department at USU. However, these datasets contained considerably less soil moisture remote sensing stations, around 95 as opposed to the 195 in the September 2017 datasets. The timeline required to obtain the additional soil moisture information fell outside the scope of this current project.

All other datasets with the 95 remote sensing sites exhibited the same problems as the September 2017 soil moisture dataset: overall poor performance from tree-based machine-learning methods using the PRISM gridded climate normals. This was the same regardless of time period observed. Tested time periods included the average of all summer month soil moisture measurements since 2016, the average of soil moisture measurements in individual months, weeks, and days. Future research will require a re-evaluation of the candidate predictor variables for soil moisture as well as methods to detect and remove data anomalies associated with irrigated sites. 

\subsection{``Smoothness'' of the maps: Characterizing a complex landscape} \label{sec:smoothness_evaluation}

Using the cross-validated RMSE score is not the only way we can evaluate the performance of these methods. In Section \ref{sec:intro} and Figure \ref{fig:rfmap}, it is mentioned that the ultimate aim of these methods is to characterize a complex landscape in a physically realistic way. 

As an example, in the field of meteorology, advection schemes are a way to model the horizontal flow of heat or matter through a certain geographic region. Advection schemes are based upon the gradients (i.e. local rates of change) which must be relatively smooth to avoid numerical precision issues in modeling. Having an extreme jump in a modeled surface can lead to an extreme local gradient, thereby disrupting the small-scale meteorology. While smoothness is not a necessary condition of predictions on all spatial datasets, most spatial and ecological datasets benefit from predictions with realistic transitions in modeled values across space. 

In a spatial or ecological setting, it may be the case that we reject a method that has a slightly higher measure of predictive accuracy in favor of a method with similar accuracy and physically realistic output. In this sense, the method that appears to be providing the map with the most detail and physical realism across the geographical landscape may be judged to be the ``superior'' method, as there is strong evidence it is characterizing the patterns inherently present in the data. We are unaware of any methods that formally balance accuracy with smoothness. In practice, this balance will be domain specific. 

A critical parameter in the formation of a regression tree is the choice of pruning: the depth that the tree is grown to. A pruned regression tree may have few terminal nodes, thus yielding a ``chunky'' landscape with sharp and discrete breaks across the space. It would unfair to say that a pruned regression tree does a poor job of characterizing a complex landscape. To circumvent this unfair comparison, in the mapping process both the simple regression tree and the autocart tree are not pruned. The stopping criteria for the autocart and regression tree growth are the same, so we are ensured that there will be a similar number of terminal nodes in each tree.

\subsubsection{2017 Utah Soil Moisture}

Although the RMSE of all machine-learning methods tested are less than impressive when compared against the ``simple'' regression discussed in Section \ref{sec:result_table}, the maps generated by the methods are promising:

\begin{figure}[H]
    \centering
    \subfloat[Regression trees]{\includegraphics[width = 0.5\textwidth]{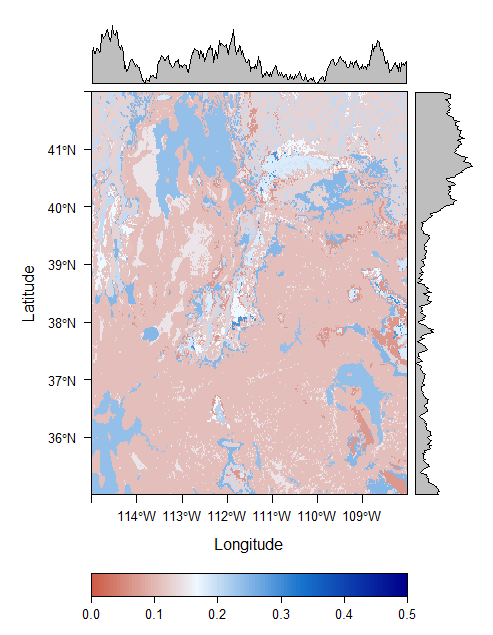}\label{fig:soil_rpart}}
    \hfill
    \subfloat[Autocart with ranged power parameter $p_1 = 0.5, p_2 = 3.0$]{\includegraphics[width = 0.5\textwidth]{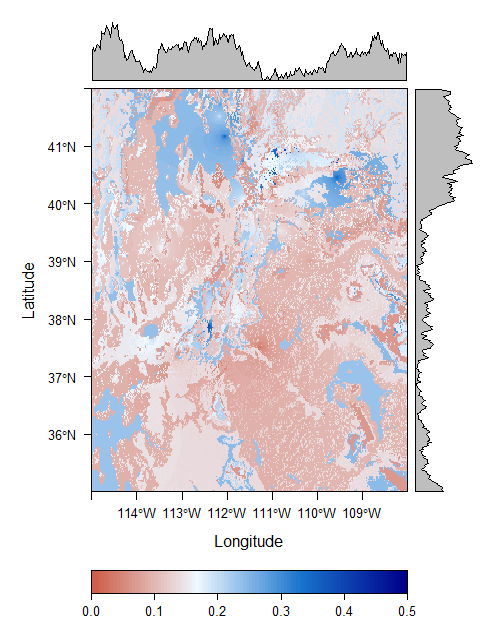}\label{fig:soil_autocart}}
    \caption{Prediction map of average soil moisture (proportion of water per $\text{cm}^3$) in September 2017}
    \label{fig:soil_map}
\end{figure}

In Figure \ref{fig:soil_autocart}, we observe the characterization of a more complex landscape, as compared to Figure \ref{fig:soil_rpart}. One noticeable ``halo'' of the interpolative effect can be seen at approximately (112 W, 41 N).

\begin{figure}[H]
    \centering
    \subfloat[Random forests]{\includegraphics[width = 0.5\textwidth]{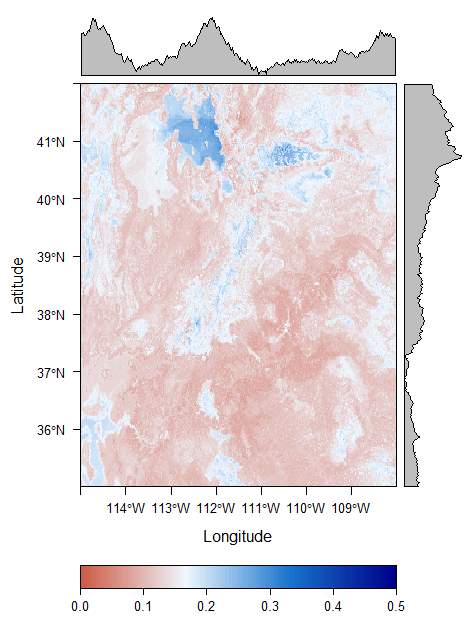}\label{fig:soil_rf}}
    \hfill
    \subfloat[Autoforest with $\text{ntree} = 100$ and $p_1 = 0.5, p_2 = 3.0$]{\includegraphics[width = 0.5\textwidth]{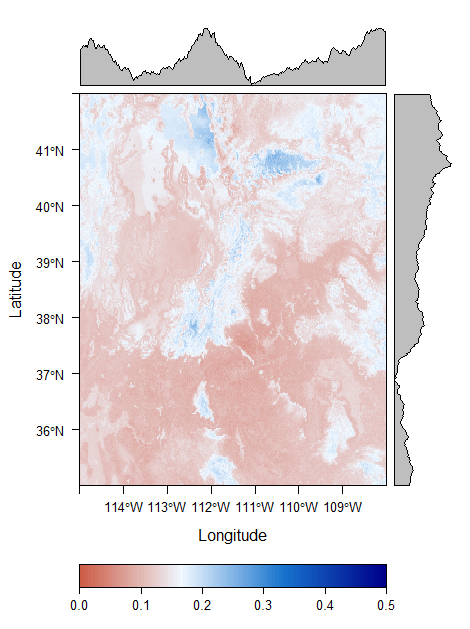}\label{fig:soil_af}}
    \caption{Prediction map of average soil moisture (proportion of water per $\text{cm}^3$) in September 2017 with ensemble methods.}
    \label{fig:soil_map_ensemble}
\end{figure}

In both the Random Forest and Autoforest maps, we see an improved characterization of the complex landscape with physically realistic predictions as compared to the traditional regression-tree based approaches. Although there is a discernible difference between the two maps, the difference is not as stark as with Figures \ref{fig:soil_rpart} and \ref{fig:soil_autocart}. 

It is important to mention that Figure \ref{fig:soil_rf} differs from Figure \ref{fig:rfmap} from the beginning of the paper in that longitude/latitude are not included as predictors in the model. As the coordinate information is not explicitly included as a predictor, we do not observe the same visual artefacts that we did in Figure \ref{fig:rfmap}.

\subsubsection{2017 Utah ground snow load}

\begin{figure}[H]
    \centering
    \subfloat[Regression trees]{\includegraphics[width = 0.5\textwidth]{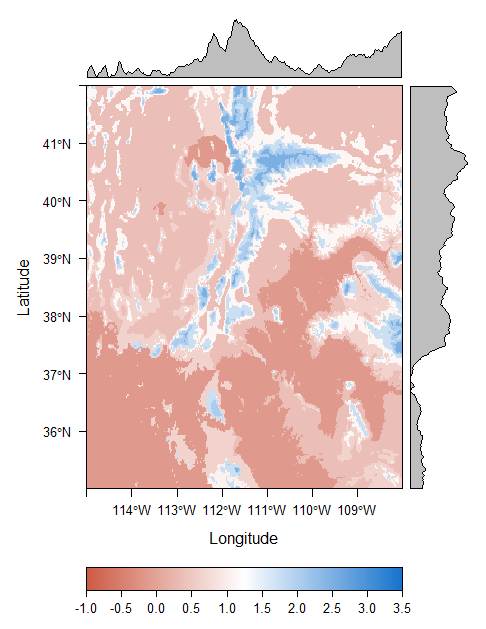}\label{fig:snow_rpart}}
    \hfill
    \subfloat[Autocart with ranged power parameter $p_1 = 0.5, p_2 = 3.0$]{\includegraphics[width = 0.5\textwidth]{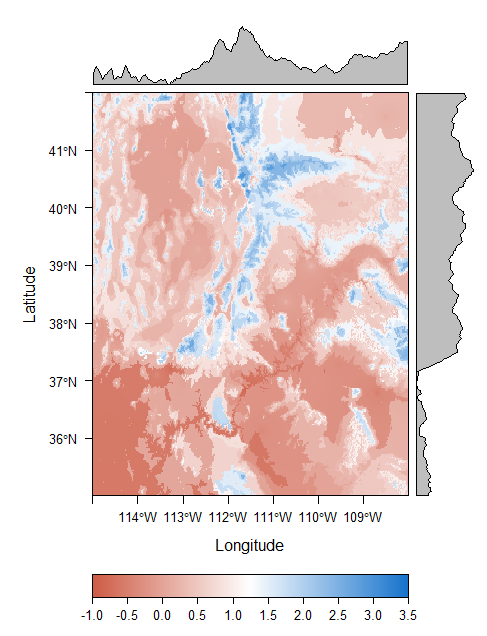}\label{fig:snow_autocart}}
    \caption{Prediction map of 50-year ground snow load average (log of kPa) ground snow load in Utah as of 2017}
\end{figure}

In a similar way to Figure \ref{fig:soil_map}, we observe that Figure \ref{fig:snow_autocart} is slightly more smooth and complex in nature. The smoothness of the autocart tree can once again be primarily attributed to the interpolation step at the terminal node of the tree.

The efficacy of both the regression tree and autocart tree is observed here, as the patterns in the map directly reflect the mountains of Utah. It is no surprise that high snow loads are observed in mountainous locations. 

\begin{figure}[H]
    \centering
    \subfloat[Random forests]{\includegraphics[width = 0.5\textwidth]{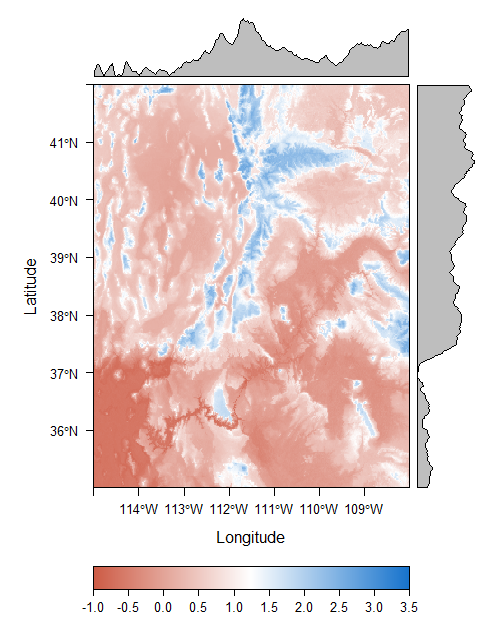}\label{fig:snow_rf}}
    \hfill
    \subfloat[Autoforest with $\text{ntree} = 100$ and $p_1 = 0.5, p_2 = 3.0$]{\includegraphics[width = 0.48\textwidth]{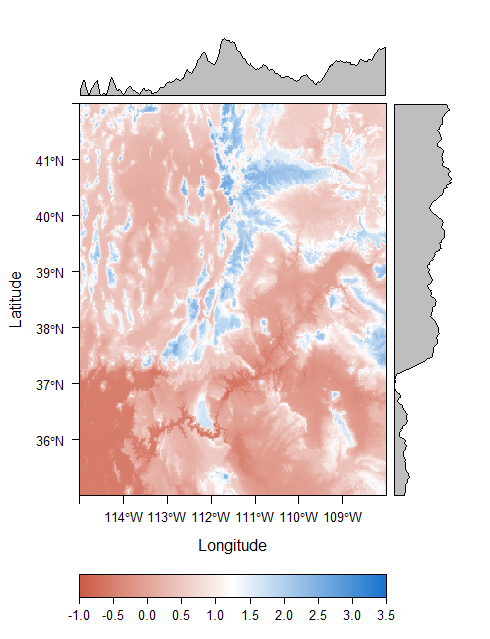}\label{fig:snow_af}}
    \caption{Prediction map of 50-year ground snow load average (log of kPa) ground snow load in Utah as of 2017 with ensemble methods}
    \label{fig:snow_map_ensemble}
\end{figure}

In the prediction map of Random Forests and Autoforest, we see very few differences. Primarily, we observe a slightly more ``patchy'' area in southwestern Utah in Figure \ref{fig:snow_rf} when compared with the smooth southwestern Utah area in Figure \ref{fig:snow_af}.

\section{Developed Software}

All software and source code surrounding the ideas in this paper are available at
\begin{center}
    \url{https://github.com/ethanancell/autocart}.
\end{center} All methods are implemented in the R statistical software environment \cite{R}. The autocart package utilizes the Rcpp \cite{package_rcpp1, package_rcpp2, package_rcpp3}, RcppArmadillo \cite{package_rcpparmadillo}, RcppParallel \cite{package_rcppparallel}, fields \cite{package_fields}, and mgcv \cite{package_mgcv1, package_mgcv2, package_mgcv3, package_mgcv4, package_mgcv5} R packages. Regression tree and random forest testing were performed using the rpart \cite{package_rpart} and randomForest \cite{package_randomforest} packages respectively.

\section{Future Work}

\begin{itemize}
    \item \textbf{Resolving possible issues in the soil moisture dataset}
    
    As mentioned in Section \ref{sec:result_discussion}, a future research direction would be in obtaining gridded predictor variables that have a stronger relationship with the measure of soil moisture. As it currently stands, the PRISM 30-year normals and other derived predictors don't appear to strongly characterize soil moisture.

    In addition, it would be helpful to have a human examination of the data so that contaminated sites (such as measurements taken on irrigated land) can be removed. The inclusion of these sites can tamper with a possible underlying relationship between predictor variables and soil moisture.
    
    \medskip
    \item \textbf{Another ensemble method of autocart trees}
    
    In addition to a ``forest'' of autocart trees, another interesting research direction would be in creating a boosted autocart trees. Boosted trees are another tree-ensemble method \cite{freund1997decision}. Boosted trees are very popular and powerful tools for both regression and classification. It may be the case that a boosted ensemble of autocart trees does comparatively better than a random forest ensemble of autocart trees.
    
    \medskip
    \item \textbf{An automatic selection of the power parameter}
    
    Wherever the ranged or unranged power parameter exists, some speculation and knowledge on the part of the researcher is required to pick a good value of $p$ or $p_1$ and $p_2$. It may be possible to include an algorithm that can automatically pick up on the weight of spatial strength in a particular region.
    
    \medskip
    \item \textbf{An objective function representing the interaction between $g_{a c}$ and $g_{s c}$}
    
    It may be the case that the objective function $g_{s c}$ described in Section \ref{sec:beta} is only useful in regions with strong autocorrelation. There may be some predictive value in selecting splits that maximize an objective function such as $\Tilde{g} = g_{a c} \cdot g_{s c} \cdot I_{\{g_{a c} > \lambda\}}$ where $I$ is the indicator function and $\lambda$ is some threshold of autocorrelation that needs to be met before the interaction objective function $g_{a c} g_{s c}$ is used in evaluating the split.
    
    \medskip
    \item \textbf{Using $g_{a c}$ and $g_{s c}$ in a non-tree-based machine-learning method}
    
    In many cases, other machine-learning methods are easily adaptable with a change in the objective function. The objective functions $g_{a c}$ and $g_{s c}$ described in this paper could be used as objective functions for other methods such as neural networks or support vector machines. 
    
    \medskip
    \item \textbf{A formal evaluation of the smoothness of a region}
    
    In Section \ref{sec:smoothness_evaluation}, the smoothness of the maps is compared with a visual assessment. The visual comparison of these maps could be supplemented with a numerical test of ``smoothness'' over a region.

\end{itemize}

\section{Conclusion}

Complex spatial datasets pose a difficult challenge to existing machine-learning methods are they are not equipped to handle spatial information in a natural way. Autocart regression trees provide a way to imbue a traditional regression tree with coordinate information in a natural way through the use of spatially-aware objective functions. The autocart tree is also capable of traditional interpolation in the terminal nodes of the tree using an adaptive inverse distance weighting technique.

Spatially-aware regression trees such as autocart also show a level of promise in providing results with a high measure of predictive accuracy on spatial datasets. In addition, the mapping result of autocart trees exhibit many desirable properties such as smoothness in the modeled response variable over the geographical region. In many cases, the mapped result of an autocart tree is very similar to that obtained by a random forest, yet retains a simpler model form.

Although the autocart method was originally created to respond to the unique challenges of modeling soil moisture in a climatically-diverse state like Utah, as of now this method does not show a particularly strong increase in predictive accuracy over a very simple regression formed by averaging the available data. It is suspected that this may be the product of a lack of soil moisture data availability, contamination of irrigation data, or an unfortunate lack of predictive power in available gridded climate variable predictors.

Due to the simple nature of the autocart tree's structure, it can be easily included in an ensemble method such as a random forest of autocart trees. Although the ``autoforest'' gain in accuracy over its traditional counterpart is not as pronounced in the random forest as it was with regression trees, it is evident that autocart trees show potential in being adapted into ensemble methods. The autocart package provides the framework and template for continued improvements in the ways that the spatial data of today is handled in the machine-learning algorithms of tomorrow. 

\bibliography{ref}

\end{document}